\documentstyle[epsfig]{mn}

\newif\ifAMStwofonts
\AMStwofontstrue

\def\be{\begin{equation}}
\def\ee{\end{equation}}

\def\gtsima{$\; \buildrel > \over \sim \;$}
\def\ltsima{$\; \buildrel < \over \sim \;$}
\def\prosima{$\; \buildrel \propto \over \sim \;$}
\def\gsim{\lower.5ex\hbox{\gtsima}}
\def\lsim{\lower.5ex\hbox{\ltsima}}
\def\simgt{\lower.5ex\hbox{\gtsima}}
\def\simlt{\lower.5ex\hbox{\ltsima}}
\def\simpr{\lower.5ex\hbox{\prosima}}

\def\ie{{\frenchspacing\it i.e. }}

\def\etal{{\it et al.~}}

\title[Chandra imaging spectroscopy of 1E 1740.7 -- 2942]{Chandra imaging
spectroscopy of 
1E 1740.7 -- 2942}  
  
\author[E. Gallo and R. P. Fender]{E. Gallo and R.P. Fender\\
Astronomical Institute ``Anton Pannekoek'', University of Amsterdam and
Center for High Energy Astrophysics, \\ Kruislaan
403, 1098 SJ, 
Amsterdam, the Netherlands\\}
\date{July 2002}

\pagerange{\pageref{firstpage}--\pageref{lastpage}}
\pubyear{2001}
\begin{document}

\maketitle
\label{firstpage}

\begin{abstract}
We have observed the black hole candidate 1E 1740.7 --2942, the
brightest persistent hard X-ray source within a few degrees of the
Galactic centre, for 10 ksec with \emph{Chandra} (ACIS-I) on August 2000.
Attempting to compensate for pile-up effects we found the spectra
were well-fit by an absorbed power law, with photon indices
$\Gamma=1.54 ^{+0.42}_{-0.37}$ (readout streak) and $\Gamma
=1.42^{+0.14}_{-0.14}$ (annulus), consistent with a black hole low/hard state.
We have 
analysed a public observation performed by \emph{Chandra} which
utilised short frames in order to avoid severe pile-up effects:
subtracting the core point spread function from the whole image, we did not
find evidence for any elongated feature perpendicular to the
radio jet axis, as reported in a recent analysis of the same 
data.
Moreover, comparing the radial 
profiles with those of an unscattered X-ray point source, we found
indication of an extended, previously undetected, X-ray scattering halo.
The measured halo fractional intensity at 3 keV is between 30
and 40 percent within 40 arcsec
but drops below detectable levels at 5 keV. 
Finally, by placing a limit on the X-ray flux from the radio emitting
lobe which has been identified as the hot spot at the end of the northern
jet of 1E 1740.7 -- 2942, we are able to constrain the magnetic energy
density in that region.
 
\end{abstract}

\begin{keywords}
binaries: general -- stars: individual (1E 1740.7 -- 2942) -- X-ray: general --
ISM: dust, extinction  
\end{keywords}

\section{Introduction}
The Black Hole Candidate (BHC) 1E 1740.7 -- 2942 is the brightest hard
X-ray source close to the centre of our Galaxy, with a hard X-ray
spectrum and luminosity comparable to Cygnus X-1 (Liang \& Nolan 1984;
Sunyaev \etal 1991a; Liang 1993). Interest in this source increased
following the discovery of its association with a double sided
radio-emitting jet by Mirabel \etal (1992), since when  it has been
classified as a \emph{``microquasar''}. Unfortunately 1E 1740.7 -- 2942
suffers from extremely high galactic extinction: intense searches,
both in the optical and IR band, have failed to identify a counterpart
(Prince \& Skinner 1991; Mereghetti \etal 1992; Djorgovski \etal 1992;
Mirabel \& Duc 1992; Marti \etal 2000; Eikenberry \etal 2001). It has
been proposed that 1E 1740.7 -- 2942 is accreting from a nearby molecular
cloud in which the source is likely to be embedded (Bally \& Leventhal
1991; Mirabel \etal 1991; Phillips \etal 1995; Yan \& Dalgarno 1997).
Recently, Smith \etal (2001) reported the detection of a weak
periodic modulation in the long-term X-ray lightcurve: a period of
about 12.5 days has been estimated.  In addition much effort has been
devoted to the possible identification of 1E 1740.7 -- 2942 with a source
of 511 keV annihilation radiation, but any definitive confirmation has
been not reported so far (Bouchet \etal 1991; Sunyaev \etal 1991b;
Anantharamaiah \etal 1993; Jung \etal 1995).  Due to its very high
hydrogen column density ($N_H \sim 10^{23}$ cm$^{-2}$ -- e.g. Sheth
\etal 1996; Churazov \etal 1996; Sakano \etal 1999) this object is in
principle an ideal candidate to investigate the properties of the
X-ray scattering halos which are known to be generated by dust grains
along the line of sight and which provide an unique opportunity to probe
the dust component of the interstellar medium (see Predehl \& Schmitt
1995) as well as potentially measuring the distance (Predehl \etal
2000).  
X-ray halos have been observed by \emph{Einstein, ROSAT} and
\emph{ASCA}, but thanks to the \emph{Chandra}'s angular and energy
resolution it is now possible to detect and investigate them in far
greater detail than ever before. More recently Cui \etal (2001)
observed 1E 1740.7 -- 2942 with \emph{Chandra} HETGS and reported the
detection of an elongated X-ray feature in the zeroth-order image, with
an extension of about 3 arcsec, orientated roughly perpendicular to
the radio-jet axis. \\
This paper is structured as follows: in section 2 we
describe the observation and present our results focusing on the
spectral analysis and on the overall structure of the source, which
has been reconstructed both from our observation and from another
public observation (see Cui \etal 2001), in which the effects of the
pile-up do not distort the central region of the image. In section 3
we show some preliminary results on the X-ray halo investigation,
comparing the radial profiles of 1E 1740.7 -- 2942 with those of an X-ray
point-like source which is supposed to be unscattered, while in the
fourth section we provide an estimate of X-ray and radio fluxes coming
from the radio emitting region which as been identified as the hot
spot at the end of the northern jet of 1E 1740.7 -- 2942. The last
section is devoted to our summary and conclusions.
%*********************************************************************8
\begin{figure*}
\epsfig{figure=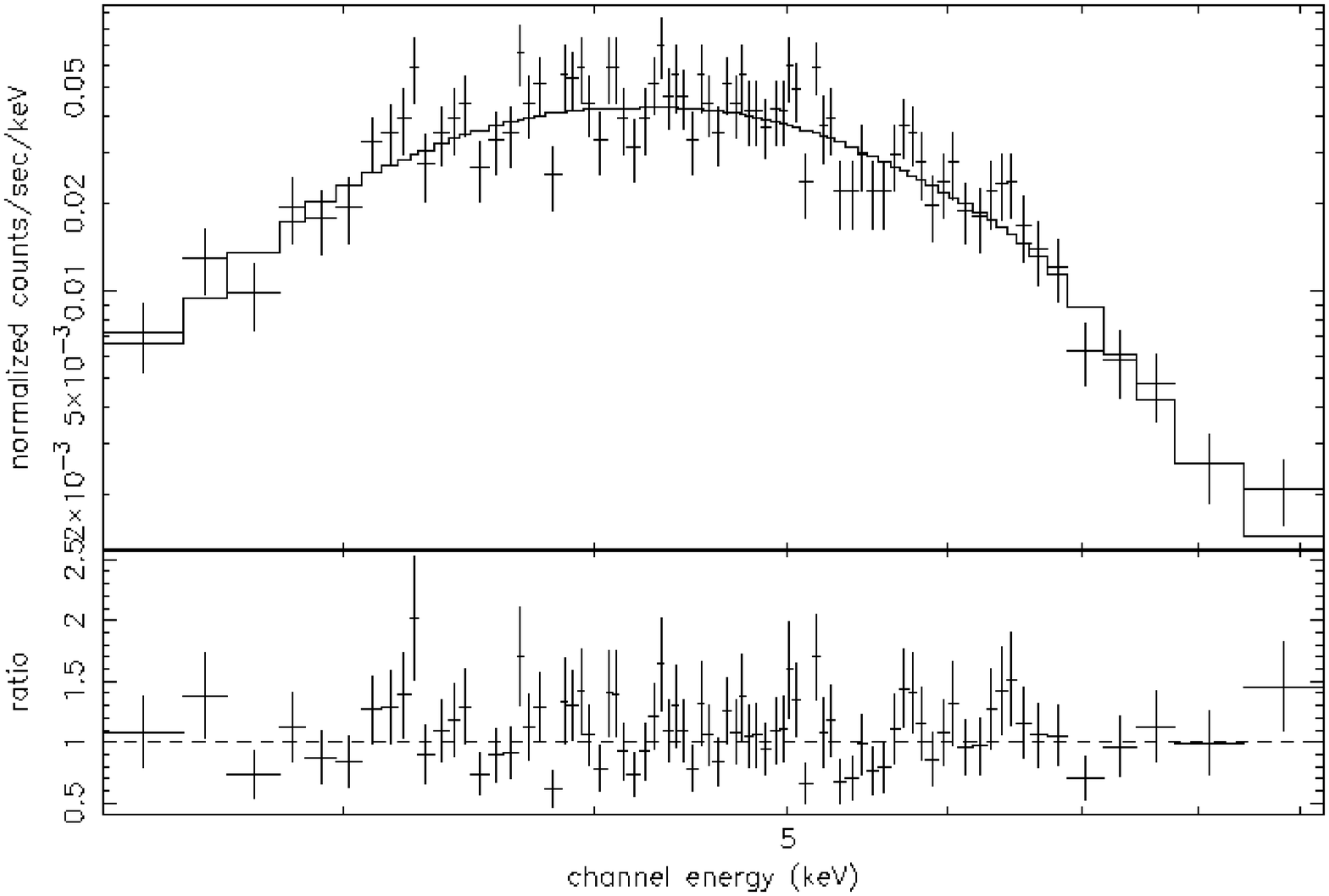,height=5.2cm}
\epsfig{figure=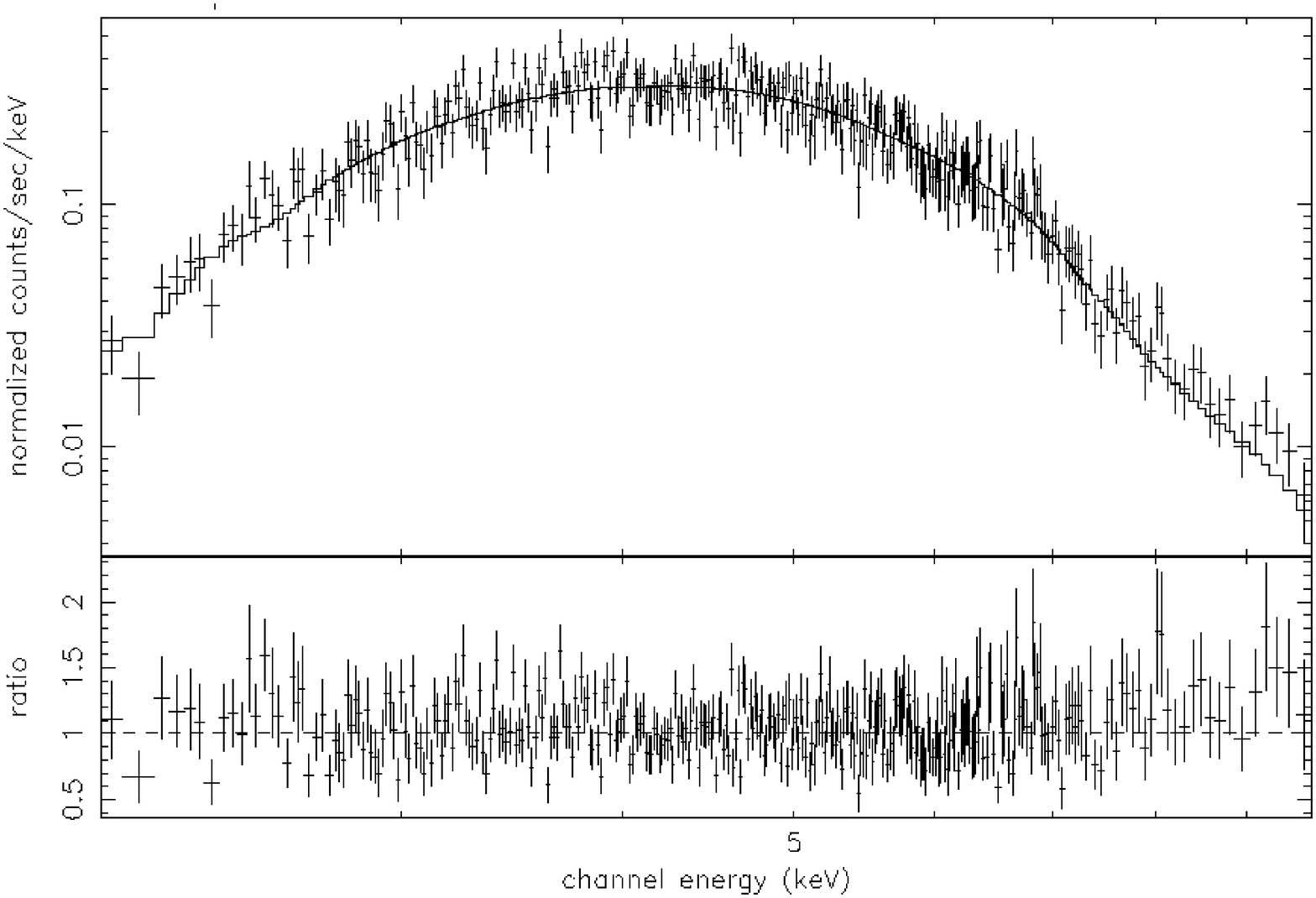,height=5.0cm}
\caption{\label{fig01}\footnotesize{\emph{Left panel}: 2-10 keV energy
spectrum of 1E 1740.7 -- 2942 extracted from the readout streak. It is
well fitted by an absorbed power law with $\Gamma=1.54
^{+0.42}_{-0.37}$, N$_{H}=11.8^{+2.3}_{-1.9} \times 10^{22}$
cm$^{-2}$} and $\chi^{2}$/d.o.f.=82/80. \emph{Right panel}: 2-10 keV
spectrum extracted from an annulus region with an internal radius of 4
arcsec. In this case we obtained $\Gamma =1.42^{+0.14}_{-0.14}$ and
$N_{H}=10.45^{+ 0.73}_{-0.70} \times 10^{22} cm^{-2}$ with
$\chi^{2}$/d.o.f.=430/331.}
\end{figure*}
%***************************************************************************

\section{Observation and data analysis}                       

We observed 1E 1740.7 -- 2942 with the \emph{Chandra} Advanced CCD
Imaging Spectrometer (ACIS, Garmire 1997) on 2000 August 30
16:59-20:00 with the nominal frame time of 3.2 sec, for a total on
source time of about 10 ks. Standard processing of the data was
performed by the \emph{Chandra} X-ray Center (CXC). The Chandra
Interactive Analysis of Observations (CIAO) tools (version 2.2)
together with XSPEC (version 11.1) have been used for analysing the
data. The ACIS-I camera consists of an array of four front-illuminated
CCDs. The physical pixel size is 0.24 $\mu$m, which at the aim-point of
the telescope is comparable to the 0''.5 spatial resolving power of
the High Resolution Mirror Assembly (HRMA). Each CCD contains 1024 x
1024 pixels organized into four readout nodes each of which reads out
1024 rows and 256 columns of pixels. The source was positioned close
to the aim-point of the telescope, which is about 960 rows away from
the readout node, on ACIS-I device I3.  As expected due to the
relatively high luminosity of 1E 1740.7 -- 2942, the image suffers from
severe pile-up effects, which affect especially the central region,
giving rise to an apparent ``hole'' of no events at the centre of the
source. 
That means that the total energy of the events is larger than
the threshold ($\sim$ 15 keV) and is then rejected. In addition
clearly visible is the ``readout streak'', which arises as a result of all
pixels  
in the column in which the bright source lies being exposed to that source for
40 microseconds. 
The measured count rate for a circular region with a radius of 4
pixels centered on 1E 1740.7 -- 2942 is 0.053 count/sec. Note that, based on
previous observations of 1E 1740.7 --  
2942, PIMMS predicts a pile-up fraction for \emph{Chandra} ACIS-I of $\sim
75\%$. 

\subsection{X-ray spectrum}

Despite the poor statistics we were able to extract the spectrum from
the readout streak which does not suffer of pile-up effects due to about 80
   
times shorter exposure. In order to do this we isolated a region of width
5 pixels (PSF FWHM is about 2 pixels), centered on the readout streak,
extending from just outside the core region to the edge of the CCD.
The
result is shown in Fig. ~\ref{fig01}, 
left panel. The best fit model ($\chi^{2}/{ d.o.f.}$=82/80)
requires an absorbed power law with a photon index $\Gamma =
1.54^{+0.42}_{-0.37}$. The hydrogen column density turns out to be
extremely high, even for a source located in the Galactic centre:
$N_{H}=11.8 ^{+2.3}_{-1.9} \times 10^{22}$ cm$^{-2}$. The integrated
unabsorbed 2-10 keV flux is $4.6 \times 10^{-11}$ erg/cm$^{2}$/sec,
corresponding to an unabsorbed soft X-ray luminosity of $4 \times
10^{34}$ erg s$^{-1}$, at a distance of 8.5 kpc.  An alternative
approach delivering more counts but possibly more vulnerable to
pile-up effects, is to take the X-ray spectrum from an annulus about
the piled-up core. In this case we have extracted the spectrum from an
annulus between 4 and 30 arcsec  (Fig. ~\ref{fig01}, right panel).
The 
best fit parameters are:  $\Gamma =1.42^{+0.14}_{-0.14}$, for $N_{H}=10.45 ^{+
0.73}_{-0.70} \times 10^{22} cm^{-2}$ with
$\chi^{2}/{ d.o.f.}=430/331$, i.e. consistent with those obtained
from the readout streak. We do not believe that scattering in the dust halo
reported in section 3 will significantly affect this spectral analysis, as the
strongest scattering will occur at energies which are heavily absorbed. 
Our analysis is consistent with a black hole low/hard state, in
agreement with Main \etal (1999), who performed a long term monitoring
of 1E 1740.7 -- 2942: the source was observed 77 times over a period 1000
days: its photon index varied between about 1.4 and 1.8. Note that
other recent papers (Sakano \etal 1999, Cui \etal 2001) have however
reported rather harder spectra than expected for the low/hard X-ray state,
with $0.9 \leq \Gamma \leq 
1.3$.

\subsection{Morphology}
\begin{figure}
\psfig{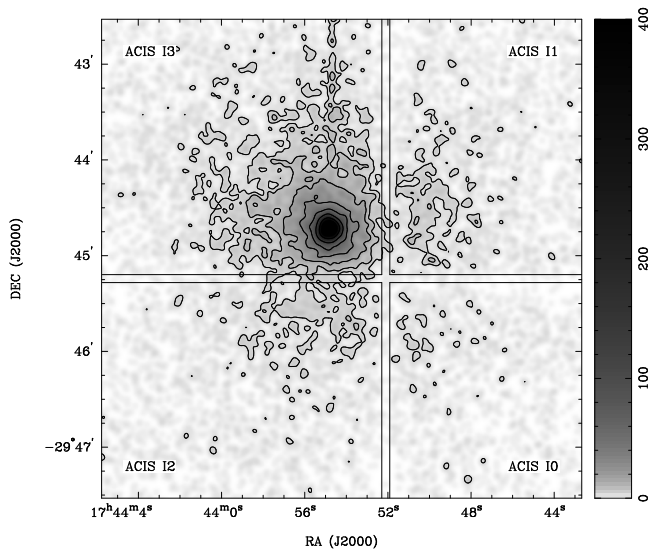}
\caption{\label{fig02}\footnotesize{Our 10 ksec ACIS-I image of 1E
1740.7 -- 2942, smoothed with a 4 arcsec Gaussian. There is no evidence for
asymmetric structures on  
arcmin-scales. The cross-like structure is due to gaps
between the chips in  
the ACIS-I array. Clearly visible is the readout trace, extending North, while
the lack of photons in the centre does not appear because of the smoothing.}} 
\end{figure}
%***********************************************************************
Cui \etal  (2001) observed 1E 1740.7 -- 2942 for about 10 ks with
\emph{Chandra} HETGS (Observation Identity 94) adopting an 
alternating exposure data mode, with three long frames (nominal 3.2 sec
readout time) 
and one short (readout time of 0.3 sec) in order to mitigate
pile-up.  Inspecting the zeroth-order image they found evidence of an
elongated X-ray feature, 
with an extent of about 3 arcsec,
roughly perpendicular to the axis of the radio lobes. In order to
establish the 
reality of this
structure they ran MARX simulations for the HETGS zeroth-order image of a
point source and fitted the simulated Point Spread Function (PSF) to the
short-frame 
radial profile, collapsed along the elongation of such a feature. Based on
this approach they estimated that the measured profile is not point like at a
confidence level of 99.9 $\%$.  \\
In
Fig. ~\ref{fig02} we show our ACIS-I image of 1E 1740.7 -- 2942, smoothed with a 4
arcsec Gaussian: we do not see any asymmetry in the X-ray
emission   
on scales of a few arcmin.
However, these data are not useful for investigating the elongated feature
reported by Cui \etal, due to the aforementioned pile-up effects. 
Nevertheless, we 
have also looked at the data of Cui \etal (2001), now publicly available, by
using a quite 
different 
approach, as follows: the CIAO package tool \emph{mkpsf} performs an
interpolation 
between  
PSF   
library files, which consist of two dimensional simulated 
monochromatic PSF images, stored in multi-dimensional FITS 
``\emph{hypercubes}''
with energies 
ranging from 0.277 keV to 8.6 keV and azimuth/elevation steps of either 1
or 5 arcminutes. We extracted the  
\emph{Chandra} PSF  
in the brightest point of the image at 5.5 keV, where the energy histogram of
the short frame image reaches its maximum.  
Then we normalised the PSF we obtained 
to the total number of counts of the source and subtracted it from the short
frame 
image of 1E 1740.7 -- 2942, in order to identify some excess.
%************************************************************************8

\begin{figure*}
\epsfig{figure=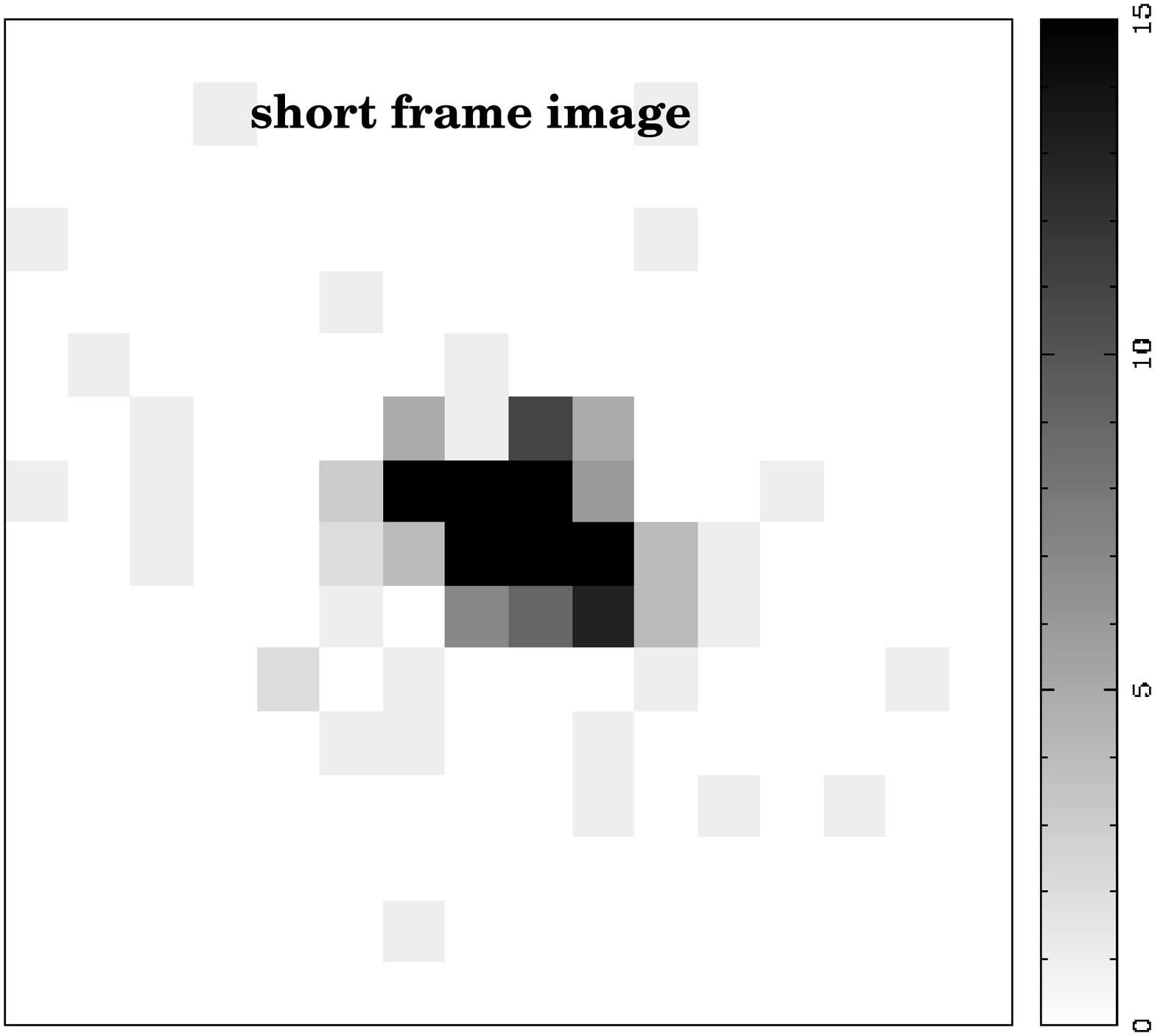,height=5cm}
\epsfig{figure=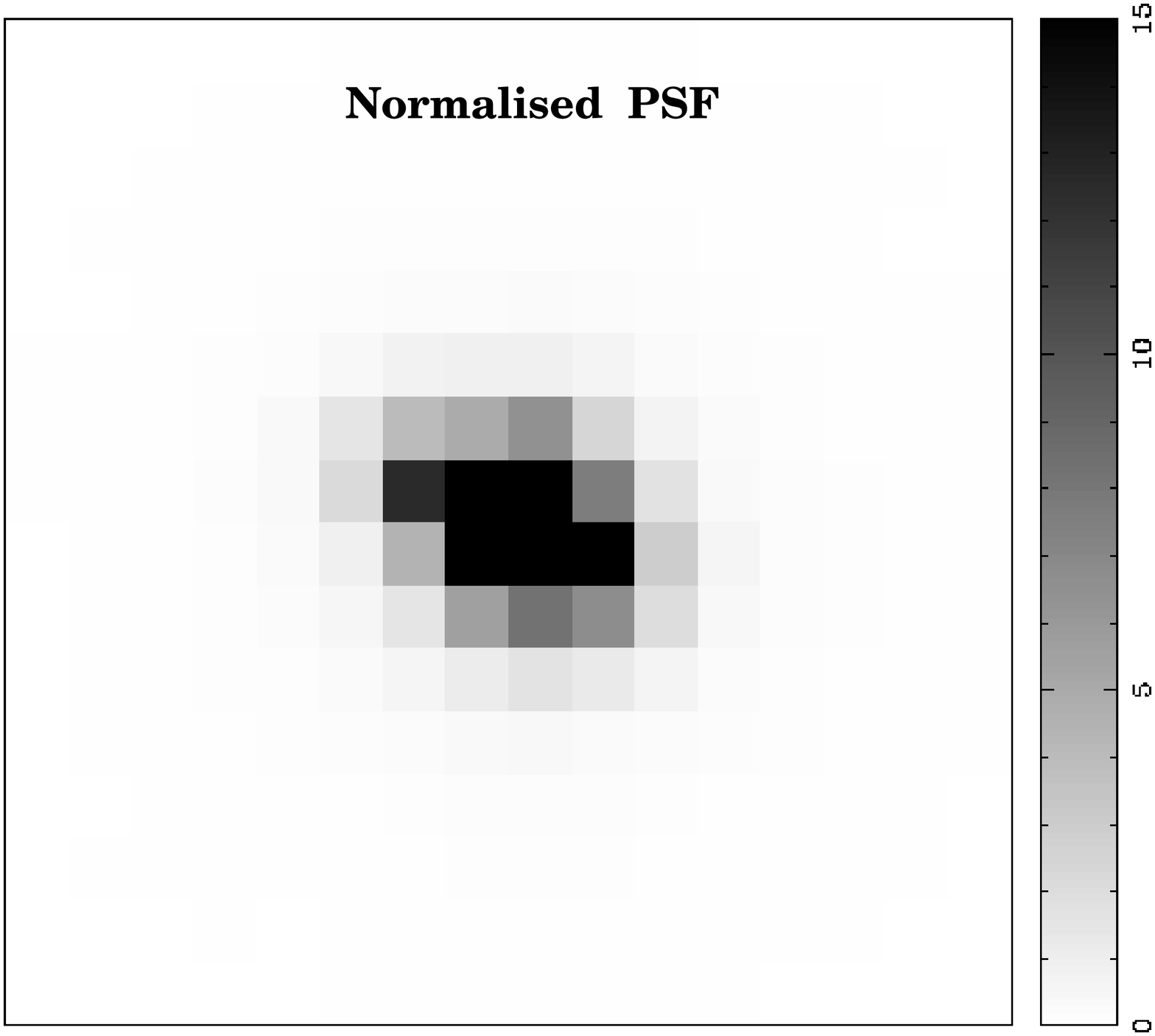,height=5cm}
\epsfig{figure=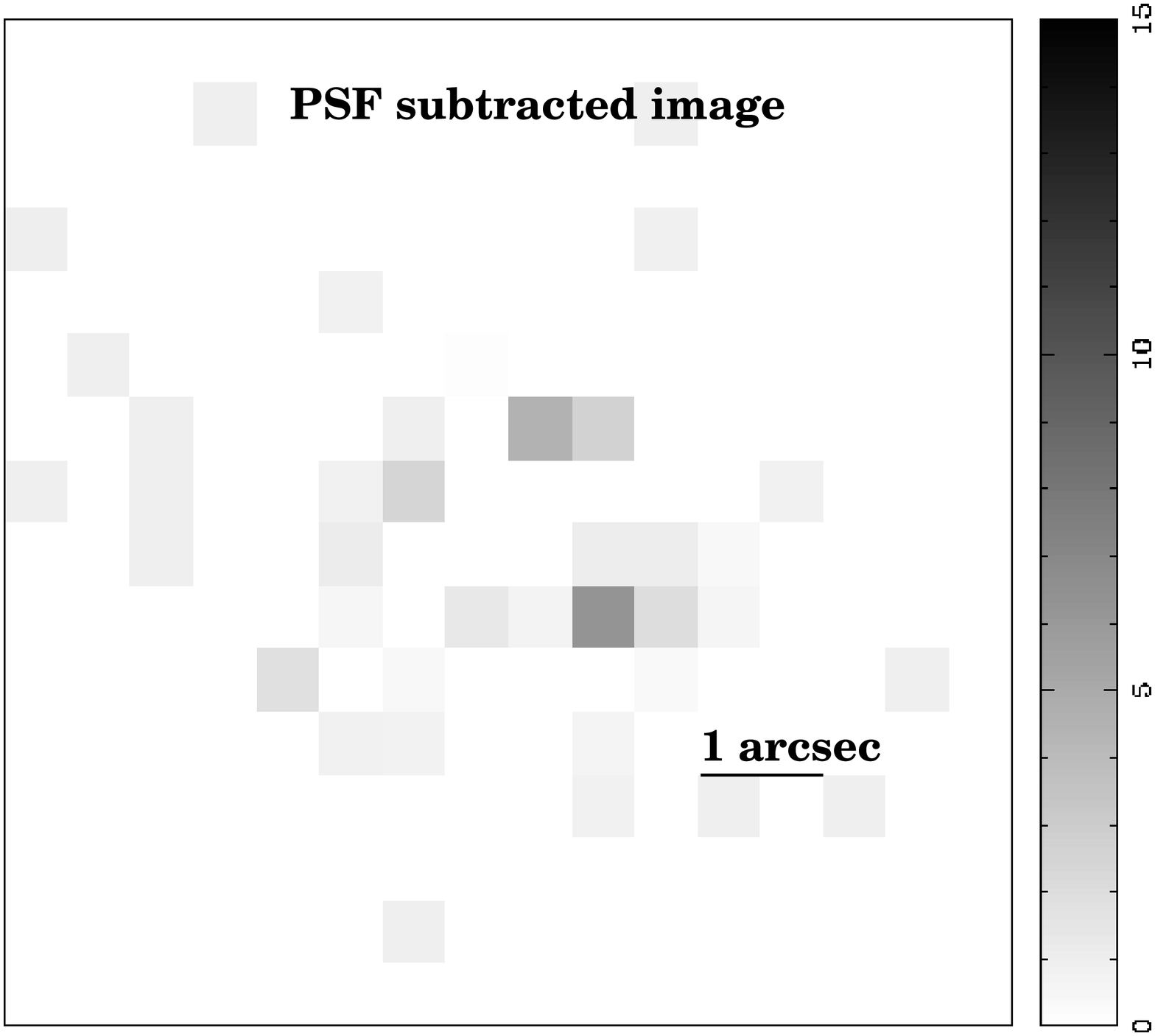,height=5cm}
\caption{\label{fig03}\footnotesize{\emph{Left panel}:
zeroth-order \emph{Chandra} HETGS observation of 1E 1740.7 -- 2942 performed in
1999 
with a short frame time (0.2 ks of net exposure, in 
order to avoid pile-up effects). \emph{Central panel}: the \emph{Chandra} PSF
we calculated through an interpolation of library PSFs. Such a calculation
has been done in order to reproduce how a point source observed by
\emph{Chandra} in the brightest point
of the previous image and normalised to the 
same number of counts should look like. \emph{Right panel}: the PSF-subtracted
short frame image. Fitting a MARX simulated PSF to the radial profile of
the short   
frame image collapsed along and across the direction of the elongation, Cui et
al. (2001) found evidence of an X-ray elongated feature roughly perpendicular
to the radio jet axis. However, comparing the same image with the right
\emph{Chandra} PSF we do not confirm such an analysis: no elongated structure
appears after the PSF subtraction from the whole image. The residuals in the
bottom panel  
are due to the fact that the PSF, which is position as well as energy
dependent,  
has been calculated at 5.5 keV, where the short frame image reaches the highest
number of counts.}}
\end{figure*}

The result is shown in Fig.  ~\ref{fig03} : simple visual inspection of
the PSF (central panel) suggests that the apparent E-W
elongation could be an artifact because it appears in
the PSF image itself.  This conclusion is confirmed by the subtraction
(right panel), which does not reveal the presence of any significant
asymmetry. The visible residuals appear to be due to the fact
that the normalised PSF, which is of course energy dependent, has been
calculated at 5.5 keV, where the short frame image shows its maximum
in the number of counts versus energy plot.  Furthermore, our
experience with observations of unscattered point sources (see below)
suggests that the genuine wings of the PSF are even broader than
those of the library PSFs.
 
\section{X-ray halo}

It is well known that sufficiently dense dust clouds along the line of
sight to a bright X-ray source are expected to produce halos of faint
and diffuse high energy emission because of the (small angle)
scattering of the primary radiation by cosmic dust grains.
Our goal was to extract the radial profiles of 1E 1740.7 -- 2942 in
different energy bands and compare them to the \emph{Chandra}'s PSF in
order to identify any X-ray excess. Following the CIAO threads we
decided to use annuli from 3-100 arcsec with 2 arcsec binning, while
an annulus from 100-110 arcsec has been used for the background
region. As first step we compared the measured profile with that of
the \emph{Chandra} PSF we obtained through an interpolation of
library PSFs, as we explained in the previous section. However, these
library PSFs have not been derived from on-orbit calibration
information; a detailed comparison to observations is still
incomplete. There are indications that the shape of the library PSFs
do not match the real data very well. In particular the wings (for
distances in excess of about 10 arcsec), which result mainly from
scattering from the microroughness on the X-ray optic surfaces, seem
to be quite different \footnote{See: \emph{
http://asc.harvard.edu/ciao/caveats/psflib.html} and
\emph{http://cxc.harvard.edu/cal/Hrma/hrma/psf/psfwings/psfwings.html}}
. Following 
the same path described in the preceding section, we generated the
\emph{Chandra} PSF, by interpolation of library files, in the centre
of our image.  Then we extracted the PSF radial profile and normalised
it to 1E 1740.7 -- 2942 at a radial distance of 4 arcsec ($\sim$ 8 pixels)
(at which radial distance pile-up effects are negligible). The slope of the
PSF radial profile generated in this way is quite steep 
compared to the data: if fitted by a simple power law model,
the profile of the PSF (in count/sec/pixel vs. pixel) has a $\Gamma=
3.2 \pm 0.1$ while the profile of 1E 1740.7 -- 2942 is well fitted by a
$\Gamma = 2.8 \pm 0.1$ power law. \\
In order to rule out the suspicion that such a discrepancy was
not due to an underestimation of the PSF wings we decided to use as
point-like source a public \emph{Chandra} observation of the high
Galactic latitude Active Galactic Nucleus PKS 2155-304 (Observation
Identity 3167).  Predehl \& Schmitt (1995), who performed a very
accurate study on X-ray halos around 25 point source observed by
\emph{ROSAT}, have identified this object as an unscattered X-ray
point source. 

\begin{figure*}
\psfig{figure=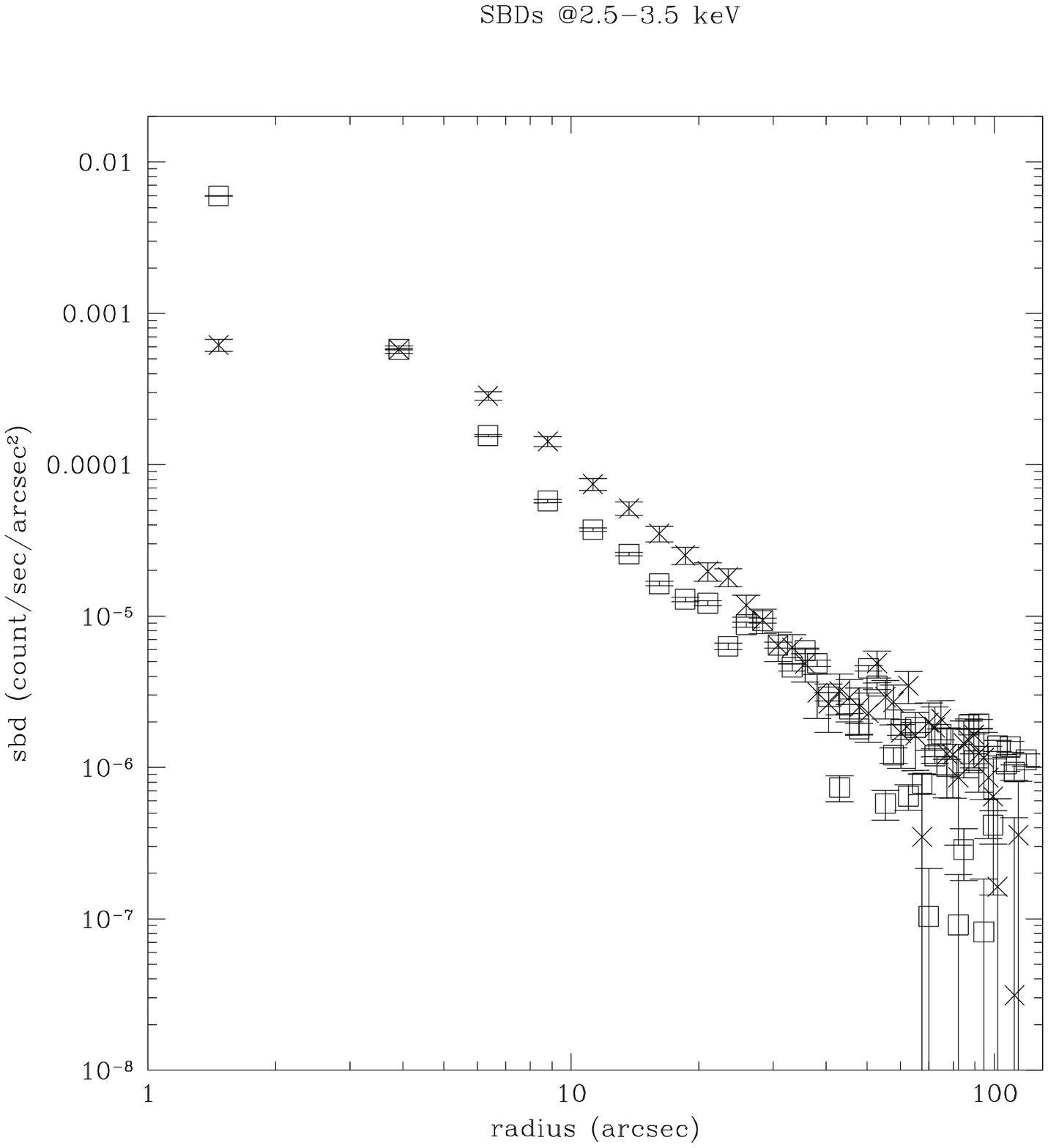,height=8.0cm}
\psfig{figure=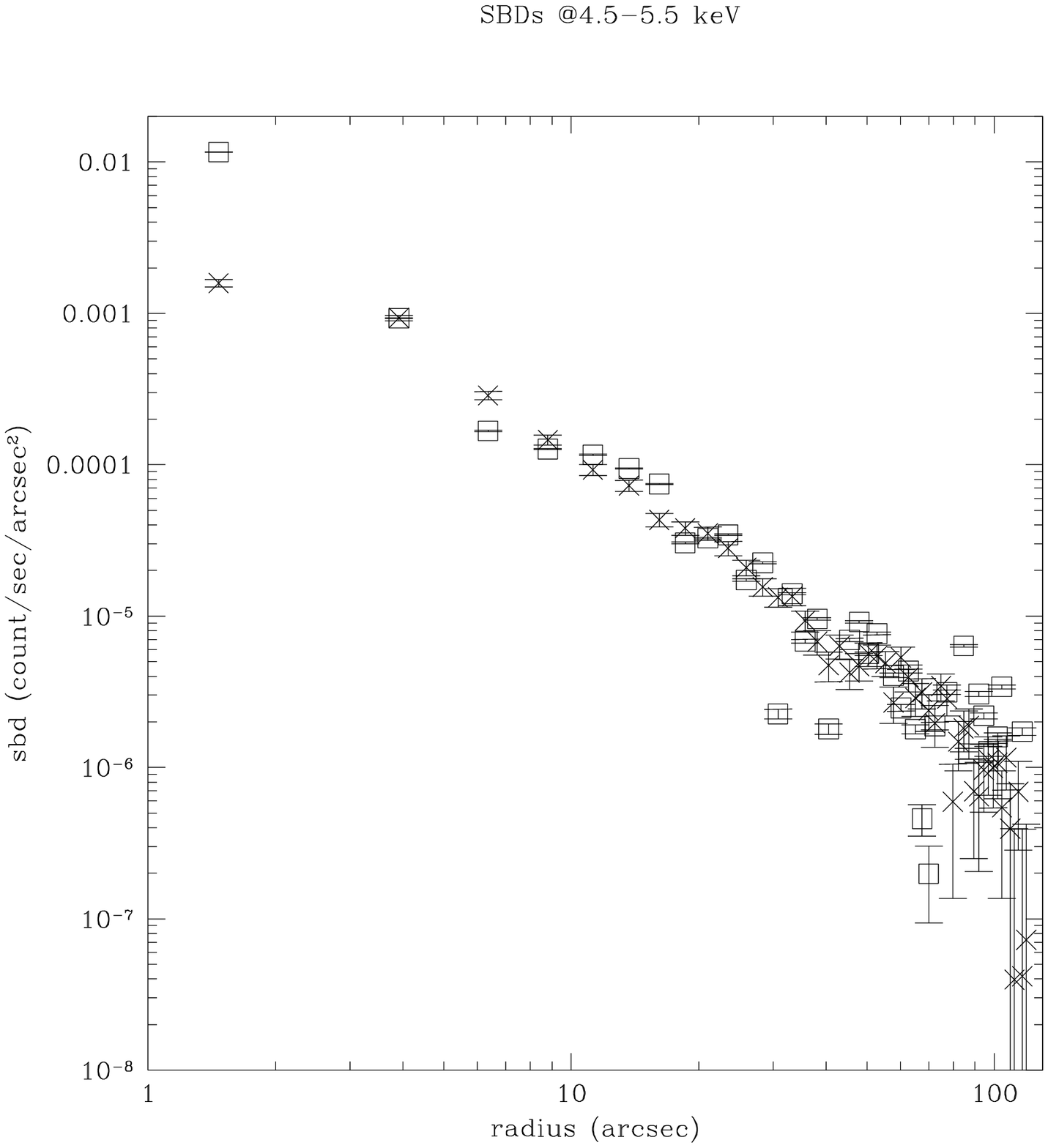,height=8.0cm}
\caption{\label{fig04}\footnotesize{Surface brightness
distributions of  
1E 1740.7 -- 2942 (crosses) 
in unit of count sec$^{-1}$arcsec$^{-2}$ vs. arcsec
compared with the unscattered X-ray point source PKS 2155-304 (squares) in two
different energy bands: 2.5-3.5 keV and 5.5-6.5 keV: left and right
panel respectively. Well recognizable are the effects of the pile-up
which  
correspond to lost and undetected events up to a radius of about 2-3
arcsec. The profiles have been normalised to the surface brightness
distribution value at about 4 arcsec, supposing that in this region the
effects of the pile-up are 
almost negligible. 
The contribution of the
X-ray halo is well recognizable in the 
softer band (bottom left panel).}}  
\end{figure*}

Our aim was to obtain the radial profile of this
source in units of count sec$^{-1}$arcsec$^{-2}$ vs. arcsec and,
again, compare it to the 1E 1740.7 -- 2942 profile.  We calculated the
surface brightness distributions using 50 annuli from 0.5-125 arcsec
(1--300 pixels).  Due to the severe pile-up effects in our observation of 1E
1740.7 -- 2942, 
the comparison between the two profiles is strongly sensitive to the
normalisation factor: as we did before, we chose to normalise them
to the measured surface brightness distribution value at about 4
arcsec, i.e. in a region where the amount of lost and undetected
events should be negligible. In Fig.~\ref{fig04} we show the results
of our halo analysis in the two representative energy bands: 2.5-3.5
keV and 4.5-5.5 keV.  Clearly distinguishable is the turn over due to
the pile-up, which completely obliterates the information in the
innermost regions for 1E 1740.7 -- 2942.
%*****************************************************************
%*****************************************************
\begin{figure*}
\psfig{figure=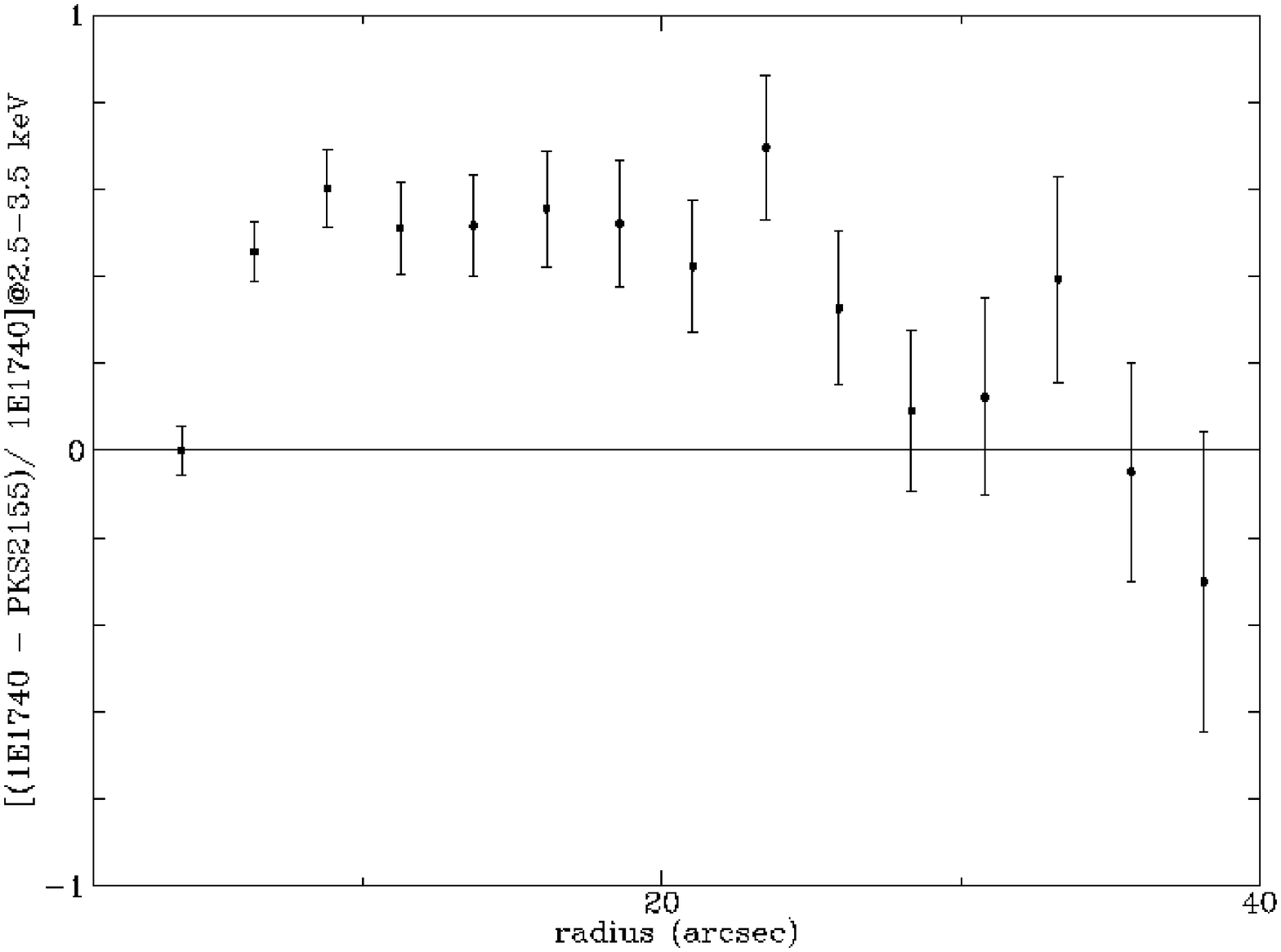,height=7.5cm,angle=-90}
\psfig{figure=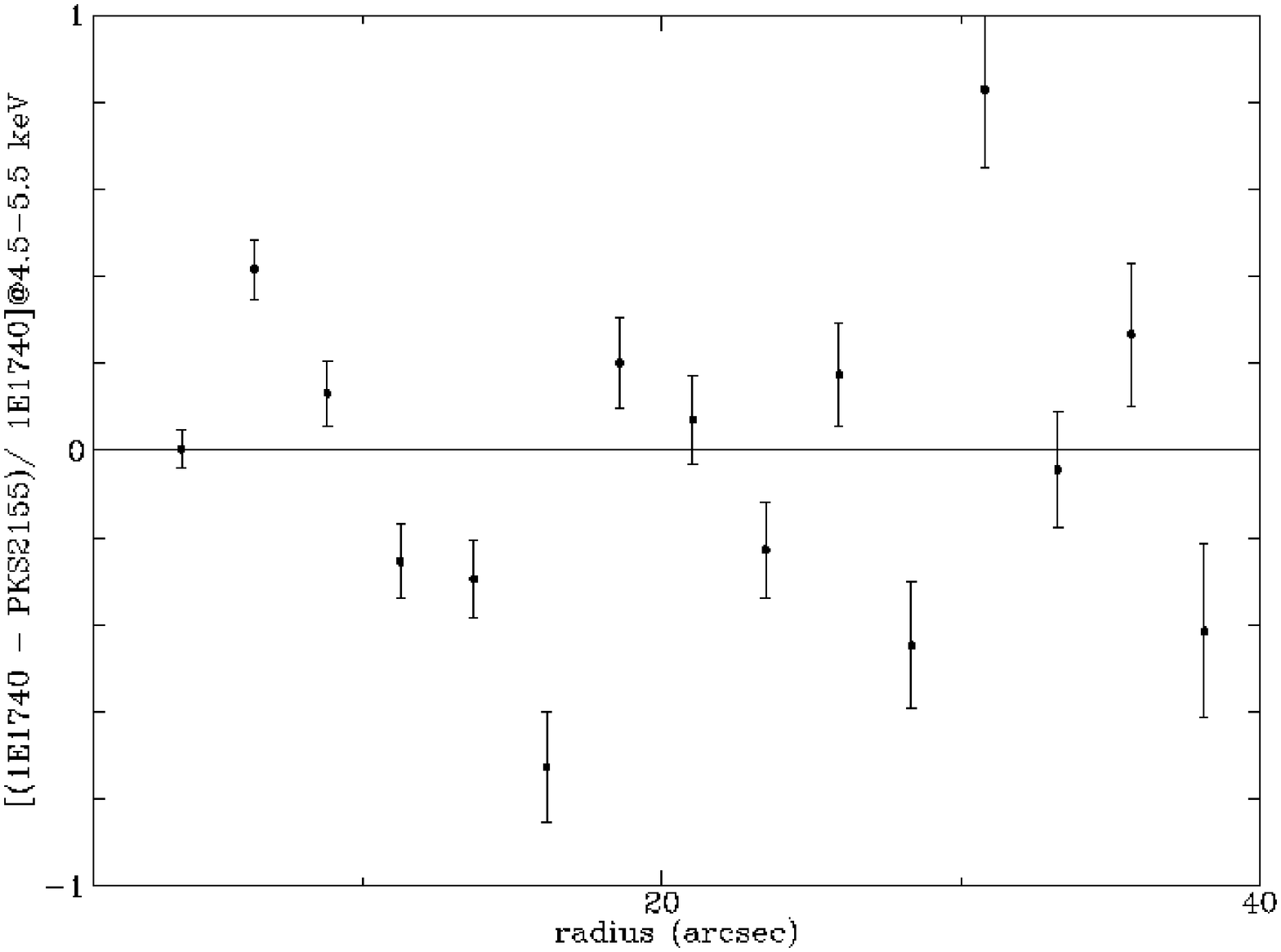,height=7.5cm,angle=-90}
\caption{\label{fig05}\footnotesize{Halo fractional intensities,
$I_{frac}$,  
\emph{vs.} radius between 2.5-3.5 keV, left panel, and 4.5-5.5 keV, right
panel. In the softer band the mean fraction of scattered X-ray photons within
40'' is
32$\pm$7 percent. In the harder band the mean fraction is $-2\pm 10 
\%$: that means that the 3$\sigma$ upper limit is only 30$\%$.  }}
\end{figure*}
%*****************************************************

Looking at the left panel, where the surface brightness distributions
between 2.5 and 3.5 keV are plotted, we can see the profile of 1E
1740.7 -- 2942 quite significantly exceeding that of PKS 2155-302 between
about 5-30 arcsec.  

The amount by which 1E 1740.7 -- 2942 exceeds PKS 2155 -- 304 is dependent on
a number of factors; seeking to minimize the effect of pile-up, as noted
above, we have 
normalised the data sets at a radius of 4 arcsec. In Fig. ~\ref{fig05} we
plot the fractional halo intensity, defined as \emph{halo/(source+halo)}, as a
function of angular distance from 1E 1740.7 -- 2942. The excess, at a level
30--40$\%$ (mean value 32$\pm 7 \%$), is detected out to 40 arcsec
from the core, in the lower (2.5-3.5 
keV) 
energy band. In the higher energy band (4.5-5.5 keV) there is no significant
excess, although the poor statistics means the 3$\sigma$ upper limit is only
30$\%$. \\
Based on Predehl and Schmitt (1995) we would expect a \emph{total} fractional
halo intensity (\ie summed over all angles) of $\sim 50\%$ at 3 keV and $\sim
20\%$ at 5 keV for such a large $N_{H}$. However, uncertainties in the angular
and energy dependence of both the pile-up and PSF, as well as in the angular
distribution of scattered photons, do not allow us to easily estimate this
value. We do note however that the apparent weakness of the halo at 5 keV may
suggest that 
most of the absorption is made locally at the source. Actually such a
possibility has already been proposed in previous works, where 1E1740.7 -- 2942
is supposed to be embedded in dense molecular cloud (Bally \& Leventhal
1991; Mirabel \etal 1991; Phillips \etal 1995; Yan \& Dalgarno 1997).
If there was
dust scattering emission potentially associated with a local evironment,
\emph{Chandra}, with 
its excellent angular resolution, should in principle be able to detect that,
although it would require a much longer observing time. \\
Obviously our analysis is biased by several approximations: first of all, the
severe pile-up effects, which make necessary the choice of a normalisation
factor between the source and PSF profiles. In addition, the choice of the PSF
itself, which, being different from the proper calibrated \emph{Chandra} PSF,
certainly introduces errors due to for instance position dependence, different
background level and so on.

\section{Constraining B in the radio lobes}

1E 1740.7 -- 2942 is associated with an unresolved, flat-spectrum radio
core and two extended, optically thin radio lobes (Mirabel et
al. 1992; Anantharamaiah et al. 1993). The high-energy electrons
present in such lobes are expected to Comptonise the ambient photons
and as a result should produce some high-energy X-ray emission.  We
have inspected our image for X-ray emission associated with the
northern lobe (lobe `B'), in order to place limits on any Comptonised
component. Unfortunately, this lobe partially lies on the gap between the
ACIS-I3 and I2 chips; however after calculating the exposure maps we
are still able to place a limit on the X-ray flux associated with the
lobe. The net count rate in a circular region of radius 8 arcsec
centred on lobe B is $2.0 \times 10^{-3}$ count/sec. We then ran a
PIMMS simulation in order to evaluate the X-ray flux emerging from
that region: we used the N$_{H}$ column density obtained from the
readout streak spectrum and chose a power law
model with a photon index of 1.5, i.e. the typical slope due to
single-scattering Comptonisation. In that way we were able to estimate
a firm upper limit to the X-ray flux, which turned out to be $\leq 8.8
\times 10^{-14}$ erg/cm$^{2}$/sec. Note that at such an angular
distance from the core, in fact, the contaminations to the X-ray
emission from the PSF and dust scattering halo are far from being
negligible; however without a good model we are not able to do any
better than this currently.  We then evaluated the radio flux between 1 and
6 GHz assuming a spectral index of $\alpha = -0.9$ (using the conventional
definition: S$_{\nu} \propto \nu^{\alpha}$), according to Mirabel
\etal 1993, and a peak flux density of 0.27 mJy at 5 GHz
(Anantharamaiah \etal 1993). In that way we obtained a radio flux of
$\sim 2 \times 10^{-16}$ erg/cm$^{2}$/sec, giving a ratio $E_{X}$/$E_{Radio} $
$\leq$ 500. This ratio constrains the ratio of photon to magnetic
energy densities, $U_{\gamma}:U_{ B}$, subject to the caveat of our
rather limited spectral coverage and assuming isotropic emission in both
bands. In the galactic centre region,  the
photon energy density has been recently estimated to be $\sim 10$ eV
cm$^{-3}$, i.e. $\sim 2 \times 10^{-11}$ erg cm$^{-3}$ (Strong,
Moskalenko \& Reimer 2000); thus our observations roughly constrain
the magnetic field in the lobes to $B \geq 1\mu$Gauss. Given that the
equipartition fields associated with the radio jets and lobes of X-ray binaries
are typically estimated to be of order mG or more (e.g. Spencer 1996), 
this is not surprising. Nevertheless, it does provide an entirely
independent lower limit on the magnetic field in the lobes.

\section{Conclusions}

We have performed a detailed analysis of a 10 ksec {\em Chandra}
ACIS-I observation of the black hole candidate 1E 1740.7 -- 2942. The
conclusions of this work are as follows:

\begin{itemize}
\item 
We have utilised two approaches to measuring the X-ray spectrum
without suffering the effects of pile-up, which are very strong in the
core of our image. Using the readout streak, the 2-10 keV spectrum is
well fitted by an absorbed power law with $\Gamma=1.54
^{+0.42}_{-0.37}$ and N$_{H}=11.8^{+2.3}_{-1.9} \times
10^{22}$cm$^{-2}$. Using an annulus from 4--30 arcsec, providing more
counts but potentially more prone to pile-up effects, we also fit an
absorbed power law, with $\Gamma =1.42^{+0.14}_{-0.14}$ and $N_{H}=10.45 ^{+ 0.73}_{-0.70}
\times 10^{22} cm^{-2}$ with
$\chi^{2}/{ d.o.f.}=430/331$. Both results are 
consistent, as expected, with a black hole low/hard state, and
furthermore indicate that the annulus approach does not suffer too
badly, if at all, from pile-up.
\item We do not confirm the presence of an elongated X-ray feature (about 3
arcsec) reported by Cui \etal 2001.  Analysing the same data set, now
public, we did not
find  
evidence for any asymmetric X-ray    
structure within about 4 arcsec from the centre. On larger angular
scales there is no obvious asymmetry -- in deeper images from our new data --
to 
several arcmin. 
\item We calculated and compared the surface brightness distributions of 1E
1740.7 -- 2942  
to those of an unscattered X-ray source in different energy bands. Despite 
complications due to pile-up effects and uncertain PSF, which lead the
comparison between 
the profiles to be very 
sensitive to the normalisation factor, we found clear evidence for scattered
X-rays in the energy range 2.5-3.5 keV at an angular separation $\lsim$ 40
arcsec 
from the core. At higher energies and/or larger angular separations, the
scattering halo is not clearly detectable.
\item We provide an upper limit on the ratio between X-ray and radio
fluxes in the region which corresponds to the hot spot at the end of the
northern radio jet emitted by 1E 1740.7 -- 2942, from which we can
crudely constrain $B_{lobe} \geq 1\mu$G.

\end{itemize}

Even though 1E 1740.7 -- 2942 is in principle an ideal candidate for the
study of X-ray scattering halos because of its huge column density,
which is proportional to the amount of dust grains along the line of
sight, such an investigation is complicated by the fact that the
source is almost completely absorbed below about 3 keV. Theoretically, the
optical depth due to scattering scales as $E^{-2}$, making the softer, 
\emph{i.e. the absorbed}, band more suitable for investigating 
X-ray halos. 
This demonstrates that a limit exists to the study of
halos which can be perfomed with \emph{Chandra} for such strongly
absorbed sources like 1E 1740.7 -- 2942, since the absorbing $N_{H}$ column is
also proportional to the amount of dust scattering. 
Comparing the halo brightness with X-ray absorption would allow to directly
quantify the amount and the density of that molecular clouds; a detailed
analysis of the relation between halo and scattering dust properties is
however beyond the aim of this paper.

Finally we would like to stress that, in order to correctly compare
the radial profile of a source which is supposed to show an X-ray
scattering halo or other extended features with that of a point-like source it
will be  
necessary a detailed comparison between simulated and measured PSF
wings profiles.
However, until a complete calibration of the PSF wings are
available, using real observations as a template remains our best option.

\section*{Acknowledgments}

We would like to thank Michiel van der Klis for useful comments and the
anonymous 
referee for his/her constructive criticism which helped significantly to
improve the 
paper.

\label{lastpage}

\newpage

\end{document}